\documentclass[usenatbib]{mn2e}
\usepackage{graphicx}

\title[Simulations for Multi-Object Spectrograph Planet Surveys]
{Simulations for Multi-Object Spectrograph Planet Surveys}
\author[S.R. Kane, D.P. Schneider, \& J. Ge]{Stephen R. Kane$^1$, Donald P.
  Schneider$^2$, Jian Ge$^1$\\
$^1$Department of Astronomy, University of Florida, 211 Bryant Space Science
Center, Gainesville, FL 32611-2055, USA\\
$^2$Department of Astronomy and Astrophyscics, Pennsylvania State University,
525 Davey Laboratory, University Park, PA 16802, USA}

\begin{document}

\maketitle

\begin{abstract}

Radial velocity surveys for extra-solar planets generally require substantial
amounts of large telescope time in order to monitor a sufficient number of
stars. Two of the aspects which can limit such surveys are the single-object
capabilities of the spectrograph, and an inefficient observing strategy for a
given observing window. In addition, the detection rate of extra-solar planets
using the radial velocity method has thus far been relatively linear with time.
With the development of various multi-object Doppler survey instruments, there
is growing potential to dramatically increase the detection rate using the
Doppler method. Several of these instruments have already begun usage in large
scale surveys for extra-solar planets, such as FLAMES on the VLT and Keck ET on
the Sloan 2.5m wide-field telescope.

In order to plan an effective observing strategy for such a program, one must
examine the expected results based on a given observing window and target
selection. We present simulations of the expected results from a generic
multi-object survey based on calculated noise models and sensitivity for the
instrument and the known distribution of exoplanetary system parameters. We
have developed code for automatically sifting and fitting the planet candidates
produced by the survey to allow for fast follow-up observations to be
conducted. The techniques presented here may be applied to a wide range of
multi-object planet surveys.

\end{abstract}


\section{Introduction}

Of all the methods used for the detection of extra-solar planets, the
radial velocity (or Doppler) technique is still the dominant source
of extra-solar planet discoveries. There are currently more than 15
independent groups searching for exoplanets using the Doppler method, most
notably the California \& Carnegie Planet Search \citep{mar96} and the High
Accuracy Radial velocity Planet Searcher (HARPS) \citep{pep04} teams, whose
combined efforts have led to the majority of exoplanet discoveries thus far.
A number of advancements in the radial velocity technique have gradually
increased the rate of exoplanet discoveries. Improvements to single-object
spectrograph design have resulted in a number of the groups obtaining
measurement accuracies close to the current practical limit for ground-based
radial velocity searches of $\sim 1 \ \mathrm{m \ s^{-1}}$. The strategy
adopted by many of the groups is to use increasingly larger telescopes which
dramatically increases (normally by at least an order of magnitude) the number
of stars which are accessible to the particular survey. Also, a few of the
groups have now been monitoring stars for a lengthy period of time such that
they are able to detect planets with orbits larger than 5~AU.

The majority of current radial velocity surveys make use of very high resolution
echelle spectrographs \citep{but96} and rely on measurements of the shift relative
to a reference spectrum, either by cross-correlation or fits to line profiles.
However, as well as being expensive, these instruments can suffer from low
throughput and can only observe a single object in each observation. A solution
to this problem is to utilise fibre-fed multi-object instruments, such as the
FLAMES instrument on the VLT \citep{pas02}. The relatively small field-of-view
(FOV) in this case is compensated by the large magnitude depth which is able to
be probed, and FLAMES has already been used to confirm several transiting
planet candidates \citep{bou04}. Another example is Hectochelle \citep{sze98}; a
high-resolution, multi-object instrument in operation on the 6.7m MMT which,
though there are currently no plans to do so, has the potential to perform an
extensive survey for extra-solar planets. Multi-object spectrographs such as
these have the potential to increase the survey volume by many orders of
magnitude and thus would have an enormous impact on the discovery rate of
exoplanets using the Doppler method.

The multi-object instrument proposed by \citet{ge02} is of a particularly unique
design which does not utilise an echelle spectrograph. A prototype of the
instrument, called Exoplanet Tracker (ET), was used by \citet{van04} to detect
the radial velocity signature of the planet orbiting 51 Peg. The Exoplanet Tracker
is a fibre-fed dispersed fixed-delay interferometer, which is essentially a
combination of a Michelson interferometer and medium resolution spectrograph. This
instrument allows the simultaneous observing of multiple targets, is relatively
cheap to build, and the lower resolution allows for a much higher throughput, thus
increasing the effective magnitude depth of the survey. The first planet
discovered using ET \citep{ge06a} demonstrated the successful application of the
interferometer design. Based on the early successes of the ET instrument using the
0.9m Coud\'e and 2.1m telescopes at KPNO, an upgraded multi-object version of the
dispersed fixed-delay interferometer, named the Keck ET, was designed for testing
on the Sloan Digital Sky Survey 2.5m telescope \citep{gun06} at Apache Point
Observatory, New Mexico. Engineering tests of the Keck ET were performed in
2005, and the first science-quality data were obtained during commissioning in
the spring and summer of 2006.

To support the variety of multi-object planetary search programs, we have
undertaken an extensive coding effort to produce simulations to guide the
observational programs. This study was originally motivated by the need to produce
accurate simulations of the data quantity and quality for surveys based upon
the observing time and target list, hence predicting the expected planet yield.
This study also provide estimates of the number of measurements required for a
significant detection and methods for reducing the number of false detections due
to noise mimicing period signals.

We present the methods and results of the simulations for multi-object planet
surveys. Section 2 gives a brief overview the development and architecture of
the simulation/fitting code. Section 3 describes the method used to generate
the simulated data including assumed planet distribution/parameters and noise
models. Section 4 explains the methods used for sifting the data for planet
candidates and fitting models to the data. Section 5 describes the expected
results from a given survey, how these results depend upon the observing
window and period distribution, and the number of planets one can expect to
transit their parent star. Section 6 presents an analysis of the number of stars
avilable as suitable candidates for a variety of fields and magnitude depths.
Section 7 discusses various possible improvements to the simulations, the
application to target selection, and stellar metallicity considerations.
Finally, it is shown how these methods and results may be applied to a wide
variety of surveys for extra-solar planets.

\section{Scientific Rationale for Multi-Object Surveys}

There are currently more than 200 known extra-solar planets. These planets
have revealed a broad range of possible exoplanetary orbital and physical
parameters which have challenged and subsequently resulted in the revision
of theories regarding planet formation. For example, the migration of
planets into short-period orbits has been explained via interactions with the
protostellar disk \citep{lin96} and other planets \citep{ras96}. Also, the
excitation of planets into highly eccentric orbits has been explained through
such processes as resonant interactions \citep{lee02} and secular interactions
\citep{ada06a} with other planets. This great diversity in planet parameters
requires a large number of new discoveries which can provide statistically
meaningful ways of understanding their origins.

Although the high-precision echelle radial velocity instruments have been
quite successful in detecting extra-solar planets, the detection rate has thus
far proceeded in an almost linear fashion due to the limitations of the echelle
techniques being used. These limitations include such aspects as the costly
approach of requiring large amounts of large telescope time and the sinlge-object
capability of the instruments, but can also include other factors such as a
relatively low throughput. In addition, many surveys have limited their target
selection to stars brighter than a visual magnitude of $\sim 8.0$; an exception
being the N2K survey which probes stars brighter than $\sim 10.5$ \citep{fis05a}.
These current echelle techniques limit the number of new planet discoveries
possible in the near future and so limit the information available to provide
refinements to planetary formation theories.

The large number statistics required to study the observed diversity amongst
exoplanets can be addressed via multi-object surveys which have the potential
to increase the detection rate by at least an order of magnitude. An advantage
of this method is that it removes the need to search only high-metallicity stars
to increase the probability of planet detection \citep{fis05b} and can therefore
yield an unbiased sample of the stellar properties of the sample stars. This
will allow for a more complete study of correlations between planet frequency
with a variety of spectral parameters. Furthermore, the distribution of planet
parameters, such as the mass distribution, will be constrained to much higher
accuracy and additional challenges to current planet formation theories possibly
revealed. A valuable bonus aspect of such a survey will be to provide accurate
absolute radial velocities for many thousands of stars, vastly increasing our
knowledge of stellar and galactic kinematics in the solar neighbourhood.

The importance of transiting giant planets to understanding planet formation is
enormous since they are the only exoplanets for which we can determine both
masses and radii. The planetary mass-radius relation is slowly developing our
knowledge of planetary structure and evolution, analogous to the stellar
mass-radius relation for stars. It is likely that a large, multi-object survey
such as the type described here will yield many transiting exoplanets through
efficient photometric follow-up of the radial velocity discoveries. These
additional transiting planets will prove invaluble to understanding the
dependence of planetary radius on the mass, age, and the composition of both
the core and the gaseous envelopes of giant planets. Further observations of
these planets using Spitzer, such as those undertaken by \citet{dem06}, will
lead to refined planetary atmosphere models which can describe in detail the
efficiency of heat transport from the dayside to the nightside of hot Jupiters
subject to large amounts of stellar irradiation.

\section{Radial Velocity Simulation Code}

Simulations for the expected planet yield play an important role in the
development of extra-solar planet surveys. There are numerous examples of
this for the transit method, such as the planet yield simulations performed
by \citet{hor03} for a variety of transit surveys and those performed
specifically for the Kepler mission by \citet{bor03}. An exhaustive FORTRAN
code for simulating transits due to exoplanets and the expected planet yield
called {\tt transim} (Transit Simulator) was developed and then applied to
the WASP0 instrument \citep{kan05}. A similar FORTRAN code, a few aspects of
which are described in this paper, has been developed for application to radial
velocity exoplanet surveys, called {\tt rvsim} (Radial Velocity Simulator).

The primary purpose of both {\tt transim} and {\tt rvsim} is to allow the user
to work with both simulated and real data simultaneously to allow accurate
comparisons to be made between them, and to produce all the fits and statistical
analysis for the data. For example, the {\tt transim} code includes a transit
detection algorithm for sifting planet candidates from the vast amount of
photometric data. The {\tt rvsim} code also includes sifting and fitting
routines which will be described in detail in later sections. Since the
observing and instrument parameters are fully customisable, the code is
perfectly able to adapt to almost any planned survey. The code was extensively
tested using realistic simulations of large-scale planet surveys. A few of the
main capabilities of {\tt rvsim} are demonstrated in this paper.

\section{Simulated Data}

To estimate how many planets we expect to detect in a given radial velocity
observing program, we performed a series of Monte-Carlo simulations which
inject planets into a realistic sample of target stars based on the known
distribution and characteristics of exoplanets. This section describes the
generation of the simulated data.

\subsection{Besan\c{c}on Models}

A realistic stellar population from which to generate our synthetic data
may be obtained via the Besan\c{c}on Galactic model \citep{rob03}. This
model is able to derive observational predictions from an overall
description of Galactic structure and evolution, where the model includes
four populations (thin disc, thick disc, spheroid, bulge). Hence, one can
simulate the Galactic stellar populations in any directions in a wide-range
of photometric bands. The resulting model produces a catalogue of
pseudo-stars for a given field from Monte-Carlo simulations of the model.
The accompanying stellar parameters are then ideal for evaluating the
feasability of observing particular fields.

For the purposes of the simulation presented in this paper, a stellar
population model was calculated for a magnitude limited survey in the
Kepler field. According to \citet{jen05}, this field covers $\sim 105$
square degrees and is centered on Galactic coordinates $l = 76.3^\circ;
b = 13.5^\circ$. \citet{jen05} used a Besan\c{c}on model to estimate that
the relocation of the Kepler field slightly decreases the number of primary
target stars, but dramatically reduces contamination due to false positive
planet detections. \citet{mah05} noted that the Kepler field is an ideal
location for a multi-object radial velocity survey since the results would
not only compliment the transit survey to be undertaken by the Kepler mission,
but also aid in target selection.

A Besan\c{c}on model tailored to the Kepler field for $6.0 < m_V < 14.0$
created a distribution of magnitudes, colours, and metallicities from which
stellar parameters were derived. Stellar radii for main sequence stars were
calculated directly from the colours provided and giant stars were excluded
from the analysis. This created a sample of 75195 stars from which 51487
giant stars were excluded, leaving 23708 main sequence stars. Figure 1 shows
the histograms of the resulting sample of stars used for the simulations.
The stellar population in the Kepler field is largely comprised of main
sequence stars with mostly solar metallicity.

\begin{figure}
  \includegraphics[angle=270,width=8.2cm]{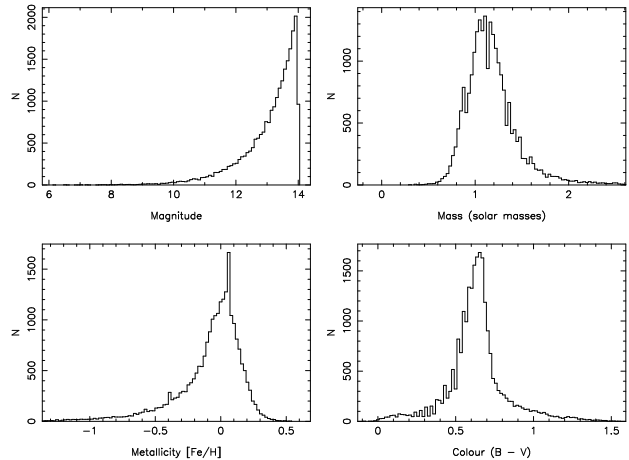}
  \caption{Properties of stars included in the simulation sample, as
    derived from the Besan\c{c}on model. Distributions include the magnitude
    (top-left), the stellar masses (top-right), the metallicities (bottem-left),
    and the $B - V$ colours (bottom-right).}
\end{figure}

\subsection{Planet Distribution}

The distribution of planets amongst main sequence stars has gradually
emerged over recent years. Several thousand nearby solar-type stars have been 
monitored by radial velocity surveys, from which it has been found that at
least 5\% of these stars harbour an extra-solar planet \citep{lin03}.
Moreover, approximately 11\% of stars which have been monitored for more
than 15 years harbour a planet and 0.5\%--1\% of solar-type stars in the
solar neighbourhood have been found to harbour a Jupiter-mass companion in a
0.05 AU (3--5 day) orbit. \citet{mar05} note that at least 6.6\% of their
survey stars harbour a planet with a mass less than 13 Jupiter masses and with
a semi-major axis within 5 AU. However, they also note that they are restricted
to M dwarfs within 10 pc due to their faintness, thus our knowledge of the
planet distribution around late-type stars is less well known.

In order to be even more quantitative regarding the planet distribution, we
can utilise the planet-metallicity correlation presented in \citet{fis05b}
which relates stellar metallicity to planetary abundance. For each of the
stars in the simulation, the probability of each star harbouring a planet was
computed based on the metallicity provided by the Besan\c{c}on model. Figure 2
shows a cumulative histogram of the planet-harbouring probability for the
sample, which demonstrates the substantial reduction in planet-harbouring stars
beyond a probability of around 5\%. As can be seen from Figure 1, the stellar
population in the Kepler field is largely comprised of main sequence stars with
mostly solar metallicity. The power law nature of the \citet{fis05b} correlation
tends to dramatically increase the number of stars with low planet-harbouring
probability for a typical metallicity distribution. In particular, stars with
metallicities less than $\sim -0.5$ dex were found to have essentially zero
probability of harbouring a planet.

\begin{figure}
  \includegraphics[angle=270,width=8.2cm]{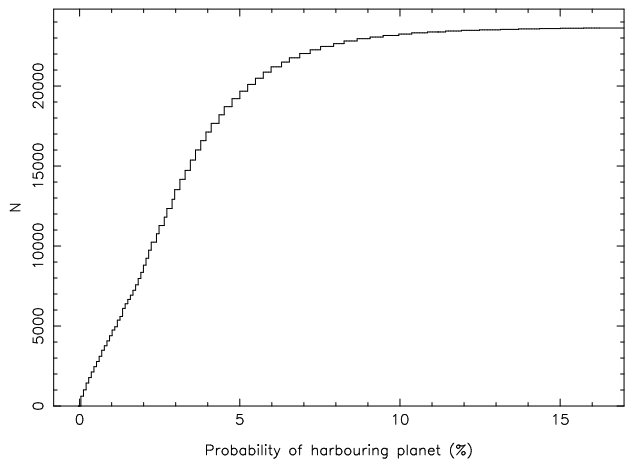}
  \caption{Cumulative histogram of the probability of harbouring a planet
for all the stars in the sample based upon the metallicity.}
\end{figure}

For the 23708 stars in this sample, a Monte-Carlo simulation of the
planet-harbouring probability yielded a result of 751 planet-harbouring
stars in the sample. Prior to this, tests were performed which assumed that
90\% of stars have planets. This initial simulation was used as a successful
test of the {\tt rvsim} code to detect planets in the data and extract the
correct planet parameters.

\subsection{Radial Velocity Calculations}

Detecting a companion to a star using the Doppler technique relies on being
able to measure a periodic change in the radial component of the stellar
velocity. The measured radial velocity, $V$, is given by
\begin{equation}
  V = V_0 + K ( \cos(\omega + f) + e \cos \omega)
\end{equation}
where $V_0$ is the systemic velocity, $K$ is semi-amplitude, $\omega$ is the
argument of periastron, $f$ is the true anomaly, and $e$ is the eccentricity.
The true anomaly, $f$, is the angle between the position at periastron and
the current position in the orbit measured at the focus of the ellipse. The
true anomaly can be expressed as $f = f(t, t_0, e)$ where $t_0$ is the time at
periastron. The semi-amplitude of the radial velocity, $K$, may be further
expressed as
\begin{equation}
  K = \left( \frac{2 \pi G}{P} \right)^{1/3}
  \frac{M_p \sin i}{(M_\star + M_p)^{2/3}} \frac{1}{\sqrt{1-e^2}}
\end{equation}
where $P$ is the period, $i$ is the inclination of the planetary orbit,
and $M_p$ and $M_\star$ are the masses of the planet and parent star
respectively. The period is also related to the semi-major axis of the
planetary orbit via Kepler's third law:
\begin{equation}
  P^2 = \frac{4 \pi^2 a^3}{G (M_\star + M_p)}
\end{equation}

Equation (1) gives the radial velocity as a function of the true anomaly, $f$.
To calculate the radial velocity as a function of time it is necessary to
find $f$ as a function of $t$. The mean anomaly is defined as
\begin{equation}
  M = \frac{2 \pi}{P} (t - t_0)
\end{equation}
and is hence the fraction of the orbital period that has elapsed since the
last passage at periastron. From the mean anomaly we can calculate the
eccentric anomaly, $E$, which is the angle between the position at periastron
and the current position in the orbit, projected onto the ellipse's
circumscribing circle perpendicularly to the major axis, measured at the
centre of the ellipse.
These two quantities are related via Kepler's equation:
\begin{equation}
  M = E - e \sin E
\end{equation}
which has the solution
\begin{equation}
  E = \frac{M - e (E \cos E - \sin E)}{1 - e \cos E}
\end{equation}
This can be solved via a Newton-Raphson iteration which converges when
$|(E_{\mathrm{new}} - E_{\mathrm{old}}) / E_{\mathrm{old}}$ is less than
some sufficiently small number. The radial velocity code in {\tt rvsim}
chooses this convergence value to be $10^{-4}$. In addition to this,
\citet{cha98} have shown that this iteration will always converge if the
initial guess of the elliptical anomaly is chosen to be $E = \pi$. This
method rapidly yields the value of $E$ (usually in only a few iterations)
and hence the value of $f$:
\begin{equation}
  \cos f = \frac{\cos E - e}{1 - e \cos E}
\end{equation}
Equation (1) can now be evaluated as a function of time.

Combining these calculations with a $\chi^2$ analysis results in a fitted
radial velocity curve from which it is possible to determine the values of
$P$, $V_0$, $K$, $\omega$, $t_0$, and $e$. The calculations are relatively
computationally inexpensive and so can be performed on a large number of
datasets in a reasonably short period of time. This is an important factor in
the methods used to search parameter space discussed in later sections.

\subsection{Planet Parameters}

Over 180 planets are now known via the various radial velocity surveys. The
number of planet discoveries and the duration of the surveys has now allowed
a more complete picture of exoplanetary parameter space. Understanding the
distribution of these exoplanetary parameters is critical to creating an
accurate simulation of the expected planet yield from a survey since this
will determine which planets the experiment is sensitive enough to detect.

As discussed earlier, each star in the simulation was assigned a
planet-harbouring probability based on the metallicity. This information
provided the basis of a Monte-Carlo simulation which showed that one could
generally expect 751 stars in the sample to harbour planets. For the
purposes of demonstrating the expected exoplanetary parameter distribution,
we have assumed that each planet-harbouring star has only one planet. Each
planet was then assigned parameters in a process repeated many times via an
additional Monte-Carlo simulation. The range of distribution of these
parameters were drawn from a variety of sources, including The Extrasolar
Planets Encyclopaedia\footnote{http://exoplanet.eu/}. In addition, more
precise estimates of planetary parameters have been recently published by
\citet{but06}. This process used here to select planet parameters will now
be described in more detail.

The first parameter to be assigned was the planetary mass. The exoplanetary
mass function has most recently been determined to be roughly described as
\begin{equation}
  \frac{dN}{dM_p} \sim M_{p}^{-1.05}
\end{equation}
by \citet{mar05}. It has been noted by \citet{jor01} and \citet{zuc01} that
this mass distribution is largely unaffected by the unknown $\sin i$
component. The inclination of the planetary orbit is chosen to randomly lie
between 0\degr and 90\degr. The discovery of a sub-Neptune mass by
\citet{bea06} implies that this mass distribution is likely to continue into
the regime of rocky planets, suggesting that terrestrial planets may be common.
By choosing a lower and upper mass limit of 0.02 and 13.0 Jupiter masses
respectively, this mass function was used to randomly determine the planet
masses. It should be noted that this does not take into account possible
variation in planet mass with host mass such as, for example, that suggested
by \citet{ada06b}. Also neglected is the apparent rise in mass with stellar
host metallicity \citep{fis05b}.

The second parameter to be assigned was the period of the planetary orbit.
Based on the work of \citet{tab02}, a good approximation of the period
distribution is to assume that exoplanetary periods are uniform in log
space. The upper and lower limits for the period distribution were chosen
to be 1 and 2000 days respectively, which is equivalent to 0.02 to $\sim 3$
AU for a solar-type stellar host. These limits encompass the majority of
the currently known exoplanets and allows for surveys durations which
extend slightly beyond 5 years. However, the smooth logarithmic function
which is assumed for this simulation does not take into account the
apparent ``period valley'' which exists between 10 and 100 days, as
noted by \citet{udr03}. There is an apparent ``pileup'' of planetary periods
which occurs near 3 days, as clearly shown by \citet{but06}. This is
accounted for in the code by suppressing periods of less than 3 days
duration to produce an identical period distribution to that seen amongst
real exoplanets.

The source of the eccentricity of extra-solar planets is generally not well
understood, particularly with regards to planet formation. Exoplanets have
been found to possess a wide range of eccentricities, mostly between 0.0 and
0.8 \citep{mar05}. These eccentricities appear to be mostly random with a
gradual increase in the eccentricity upper limit with increasing semi-major
axis. The exception is planets with a semi-major axis less than 0.1~AU, where
the orbits have been forced into nearly circular orbits due to tidal
circularization. Therefore for this simulation the eccentricity distribution
was approximated by allowing it to be randomly selected between 0.0 and 0.8.
For those orbits inside of 0.1~AU, the eccentricity was randomly selected
between 0.0 and 0.1.

With the simulated planet parameters described above, the amplitude of the
induced radial velocity was calculated for each star using equation (2).
This process of parameter simulation and radial velocity calculation was
repeated many times through a Monte-carlo simulation, the results of which
are shown in Figure 3. Based on the radial velocity amplitude histogram,
the number of expected planet detection can then be estimated based on the
precision of the experiment.

\begin{figure}
  \includegraphics[angle=270,width=8.2cm]{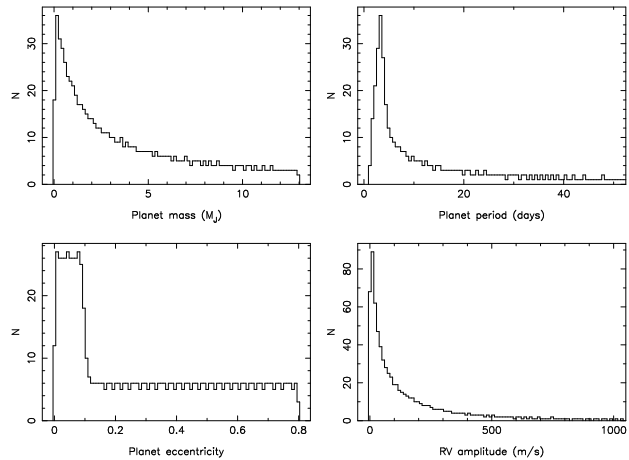}
  \caption{Results of a Monte-Carlo simulation of the expected distribution of
planet parameters. Distributions include the planet mass (top-left), the
planet period (top-right), the planet eccentricity (bottom-left),
and the radial velocity amplitude of the parent star due to the planet
(bottom-right).}
\end{figure}

\subsection{Noise Model and Error Sources}

To approximate error bars for the simulated radial velocity data, one must
consider the various sources of noise associated with the data.
The standard noise model takes into account detector characteristics as well
as photon statistics and takes the form
\begin{equation}
  \sigma^2 = \sigma_0^2 + \frac{(f_\star + f_\mathrm{sky}) \Delta t}{G}
\end{equation}
where $\sigma_0$ and $G$ are the CCD readout noise (ADU) and gain
(e$^-$/ADU) respectively, $f_\star$ and $f_\mathrm{sky}$ are the
star and sky fluxes respectively, and $\Delta t$ is the exposure time.
For the purposes of this particular application, it is sufficient to consider
the various components of equation (9) individually and then combine them
together.

There are three components which contribute to the error in the observed
radial velocity. The first is the photon noise, $\sigma_p$ which is
approximated as a simple scaling law
\begin{equation}
  \sigma_p = \sqrt{100^{(m_V - 8.0)/5}} \times \sigma_{8.0}
\end{equation}
where $\sigma_{8.0}$ is the rms photon noise at a $V$ magnitude of 8.0. The
photon noise is generally the most important of the sources of error since it
effectively defines the depth of the survey. For this simulation, we adopt a
value of $\sigma_{8.0} = 2.8 \ \mathrm{m \ s^{-1}}$ for a one hour exposure
based on typical photon noise estimates from various surveys, studies
performed by \citet{bou01}, and estimates provided by \citet{ge06b}.

The second component is the noise contribution due to systematic errors,
$\sigma_s$. Noise contributions due to systematic errors are often the most
frustrating component since they can originate from a wide range of sources
including the hardware and the data reduction software. These include
imperfections in the grating, biases inherent in the CCD detector,
flat-fielding, and moon illumination. A multi-object instrument may have
additional sources of systematic errors such as cross-contamination between
spectra, aberration/distortion of the spectra near the edge of the detector,
and guiding errors from tracking multiple stars. Based on conservative
estimates from these various noise sources, a systematic rms error of
$\sigma_s = 3.0 \ \mathrm{m \ s^{-1}}$ is assumed for this simulation.

This third source of errors considered here is that resulting from stellar
intrinsic variability caused by chromospheric activity, denoted as $\sigma_i$.
The causes of stellar intrinsic radial velocity noise has been described in
great detail in the literature, for example by \citet{saa98} and \citet{saa00}.
The recent revision of planet parameters by \citet{but06} allowed a more robust
quantitative estimate of this ``jitter'' noise by employing the methods
described in \citet{wri05}. We account for the intrinsic variability of our
simulated stars by adopting $\sigma_i = 3.0 \ \mathrm{m \ s^{-1}}$, which is
the estimated median jitter for G and K dwarfs. It should be noted that this
does not include stellar seismic activity (p-modes) \cite{dal06}. This activity
is well illustrated by the study of the $\mu$ Arae system \citep{san04,bou05}.
The high frequency of seismic activity often means that it only significantly
effects observing programs which use a relatively short exposure time, such as
HARPS \citep{mos05}.

\begin{figure}
  \includegraphics[angle=270,width=8.2cm]{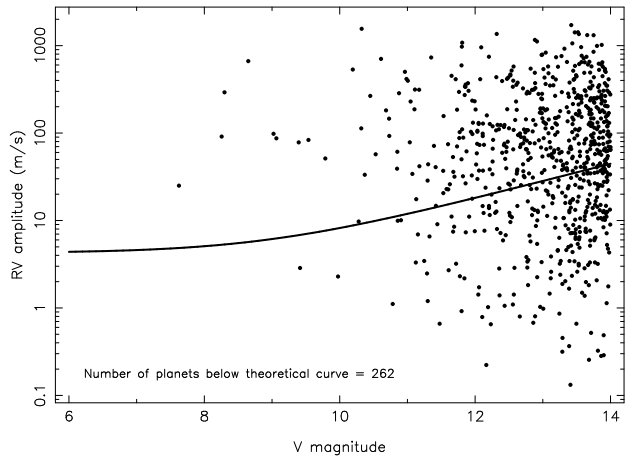}
  \caption{A radial velocity amplitude vs magnitude diagram for the simulated
    planets, showing the theoretical accuracy curve based on the noise model
    of the experiment.}
\end{figure}

The three components are then added in quadrature to obtain the total rms
error per star:
\begin{equation}
  \sigma^2 = \sigma_p^2 + \sigma_s^2 + \sigma_i^2
\end{equation}
Figure 4 shows a typical distribution of the radial velocity amplitude of
simulated planets over the full magnitude range. Also shown is the theoretical
accuracy curve (equivalent to the $1\sigma$ threshold using the above formula);
the number of planets that fall below this curve are therefore unlikely to be
detected with the current experiment.

This simple model is not without limitations but is more than sufficient for the
simulation presented here. For example, purely photon noise scaling is just
indicative for fainter stars and it is likely that a systematic noise threshold
is reached for lower fluxes. In addition, the intrinsic stellar variability does
vary with spectral type and luminosity class, particularly for M dwarfs and
subgiants, which means the jitter will be underestimated slightly in some cases.
The simulations presented here only include a single planet for each
planet-harbouring star, whereas there are likely to be additional planets in
longer period orbits whose low amplitude signal will initally contribute to the
jitter noise. Further observations will extract the periodic nature of these
signals once there has been sufficient phase coverage of the planetary orbit.
Spectroscopic binary stars are highly likely to be present in the real planet
survey data but are not considered here since they are easily detected and
readily distinguishable from planetary candidates.

\subsection{Radial Velocity Precision and Planet Detectability}

An important quantity to determine from the simulation is the number of planets
detectable for a given instrument precision. Recall that from the Besan\c{c}on
model of 23708 stars, a Monte-Carlo simulation of the planet-harbouring
probability showed that $\sim 750$ of those stars would harbour a planet. The
distribution of known planet parameters was then used to calculate the expected
distribution of radial velocity amplitudes for those stars. This information can
then be used to determine approximately the number of planets one can detect as
a function of the rms radial velocity precision of the instrument, assuming that
the precision includes noise from all sources.

Figure 5 shows the number of planets detectable from the simulated data at the
$3\sigma$ level as a function of the rms radial velocity precision. The range
of precision values shown was chosen to encompass a broad range of radial
velocity experiments. As shown in the figure, for a radial velocity precision
of $10 \ \mathrm{m \ s^{-1}}$, one may expect to detect $\sim 520$ of the
planets in the simulated sample. Likewise, for a radial velocity precision
of $5 \ \mathrm{m \ s^{-1}}$, one may expect to detect $\sim 601$ of the
simulated planets. In this example, doubling the precision of the instrument
(from $10 \ \mathrm{m \ s^{-1}}$ to $5 \ \mathrm{m \ s^{-1}}$) only increases
the planet yield by $\sim 15$\%. In general, doubling the number of survey
stars will obviously increase the planet yield by 100\%. Thus, the design of
the experiment will depend upon whether one aims to increase mass sensitivity
or increase overall planet sensitivity, since improving the radial velocity
precision is not nearly as important as increasing the number of survey stars
for increasing the net planet yield. This demonstrates the power of new
multi-object planet surveys for dramatically increasing the number over known
exoplanets over a relatively short period of time. However, this calculation
does not take into account that an improved radial velocity precision would
allow fainter stars to be available for follow-up observations, assuming that
the experimental accuracy is photon-limited. Another important point is that
this calculation only takes into account single measurements, whereas multiple
measurements will of course improve the rms precision at which one is able to
detect signatures of a given amplitude. In particular, folding the data over
multiple periods will significantly increase the power of the planetary signal
and therefore improve the detectability of the planet. This is true for both
single and multiple object surveys.

\begin{figure}
  \includegraphics[angle=270,width=8.2cm]{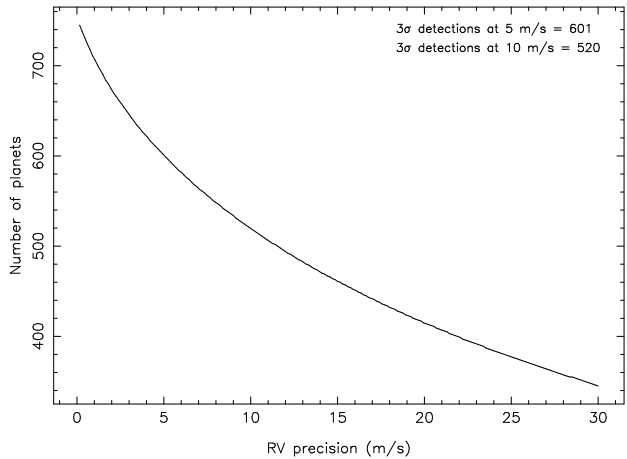}
  \caption{The number of planets detectable at the $3\sigma$ level from the
    simulation sample as a function of the rms radial velocity precision of the
    experiment.}
\end{figure}

\subsection{Observing Window and Interruptions}

The duration of the observing window is of key importance since it determines
the orbital periods to which one is sensitive from inital observations of the
targets. For the simulated data, an observing window of 30 days was assumed
which allows a broad range of tests to be performed which can be scaled to
the size of the real observing window. It is also assumed that a single
observation is obtained per star per night. These times of observations were
passed through a gaussian filter, depending upon the number of observations
per night per star, which ensured that each star was not observed at exactly
the same time each night.

However, any observing run is subject to the unpredictable nature of the
weather. This is taken into account by specifying the fractional number of
clear nights for a given observing site. For each simulated star, it is randomly
determined per night if data was acquired on that target. For this simulation,
we have assumed that 60\% of the nights are available for spectroscopic
observations, which is typical for most observatories.

\section{Radial Velocity Fitting}

In this section we describe the method used for automatically sifting planet
candidates from the data and the iterative grid-search technique used to fit
models to those candidates. We also discuss the false-alarm rate and the
dependence of successful detections upon various star and planet parameters.

\subsection{Sifting the Data}

The formidable task that remains once the data has been acquired and processed
is to efficiently extract planet candidates from the dataset. This is
particularly important for large-scale surveys, transiting planet surveys
for example, in which a relatively automated method must be constructed in
order to (a) keep the data analysis rate at a level which is at least as fast
as the data acquisition rate, and (b) to prevent missed detections since
sifting the data manually will not always be able to discern a periodic
signature unless phase-folded. The {\tt rvsim} code will ingest either real
or simulated data and attempt to perform such a sifting operation by making
use of data statistics combined with a weighted Lomb-Scargle fourier period
analysis.

For the period analysis, we make use of the Lomb-Scargle (hereafter L-S)
periodogram method \citep{lom76,sca82} which is especially suited to unevenly
sampled data. This method, which uses the Nyquist frequency to perform spectral
analysis of the data, produces a ``L-S statistic'', which is the fourier power
calculated at each frequency over a range of frequencies. The maxmimum fourier
power is the frequency which yields the least squares fit of a sinusoidal model
to the data. The L-S statistic also indicates the significance level of the fit
at that particular frequency and hence reveals the likely value for the period.
If one scans over $M$ independent frequencies, then for a fourier power $P(f)$
at frequency $f$, the probability $P$ that other frequencies in the scanned
region have a higher power is
\begin{equation}
P (> P(f)) = 1 - (1 - e^{-{P(f)}})^M
\end{equation}
which is also referred to as the false-alarm probability. For example, a
probability of 0.1 indicates that the peak in question has a 10\% chance of
being due to a false-alarm and thus we are 90\% confident in the significance
of the peak at that frequency. Figure 6 displays an example periodogram for one
of the simulated stars. The dotted lines show the false-alarm probabilities
from 50\% to 0.1\%.

\begin{figure}
  \includegraphics[angle=270,width=8.2cm]{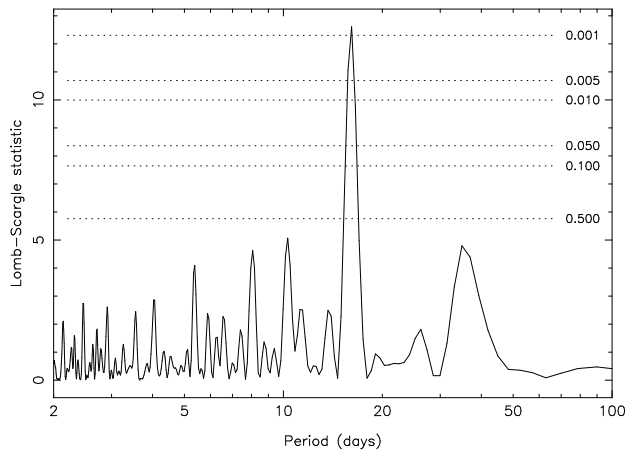}
  \caption{An example L-S periodogram for a simulated star which harbours a
  planet. The dotted lines indicate the probability that the fourier power is
  a false-alarm and hence the confidence one can have in the period.}
\end{figure}

An issue which can arise when using the standard L-S periodogram is the fact
that experimental uncertainties are neglected entirely in the calculation of
the fourier power. The standard L-S is inversely proportional to the variance
of the entire dataset which is completely independent of the individual
variances on each data point. In general, the standard L-S provides a robust
period analysis assuming that the error bars are reasonably similar such as
data acquired from the same instrument. In cases where one wishes to combine
data from several sources which may have substantially different uncertainties,
a more complete period analysis can be achieved by appropriately weighting the
fourier power calculation. An example of this can be seen in \citet{ge06a} where
radial velocity data from KPNO was combined with follow-up data from the HET.

A weighted L-S period analysis for {\tt rvsim} was created using the methodology
shown in \citet{aha05}. This method applies inverse variance weighting
calculated for each data point to each of the terms in the fourier component. In
cases where the error bars (and hence the weights) are equal, the weighted L-S
periodogram reduces to the standard L-S periodogram.

The first step of the automated fitting routines of {\tt rvsim} is to subject
each dataset to a number of statistical tests that categorise the star based
on the outcome of the tests. The first test is a weighted L-S period analysis
which attempts to detect a periodic signature in the data. If a frequency is
found for which the false-alarm probability is less than a user specified value,
50\% for example, then the data is classified as ``periodic'' and is passed to a
fitting code (described later) for more thorough analysis. It should be noted
that periods of less than $\sim 1$ day are assumed to be aliases and are
discarded in favour of the next highest fourier peak. This not only leads to a
more accurate period estimation but also dramatically decreases the rate of
false detections.

If, on the other hand, no periodic signal is detected then a variability
statistic based on the rms scatter and standard deviation of the data is used
flag those stars for which there is sufficient variability. These stars are
classified as ``variable'' and are noted for future investigations since planets
with periods substantially longer than the observing window are unlikely to show
a periodic signature until more phase coverage is achieved. If a star fails to
be classified as either periodic or variable then it is classified as
``unknown'', which means it is presumed to be a constant star at least until
further data can be acquired. As shown in the following sections, this technique
is an effective method for recovering planetary signatures in radial velocity
data down to the photon noise-limit of the instrument.

\subsection{Detection Efficiency and False-Alarm Rate}

The efficiency of this method for detecting real planetary signatures in the
data and its frequency of false detections is dependant upon a number of
factors. The most obvious and dominant factor is the false-alarm probability
threshold which one chooses to classify data as being periodic. As the
probability threshold is increased, the number of successful planet detections
will increase but the number of false detections will also increase. Although
one seeks to keep the number of missed planetary detections to a minimum, the
issue of false detections can become serious if the required follow-up
observations begin to substantially compromise the total observing time of the
experiment and thus compromise the overall planet yield.

In order to determine exactly how these detections depend upon the
false-alarm probability threshold, the method was tested using a specific
dataset and a range of probability threshold values. Shown in Figure 7 is the
resulting relation on the probability threshold for false detections, real
detections, and missed detections. The figure shows that the number of false
detections is sensitively dependant on the false-alarm probability threshold,
whereas the number of real detections is relatively independent of the cutoff.
If one chooses to restrict the periodic signatures to those which pass a 10\%
probability threshold, one will have eliminated essentially all of the false
detections without severely compromising the total planet yield.

\begin{figure}
  \includegraphics[angle=270,width=8.2cm]{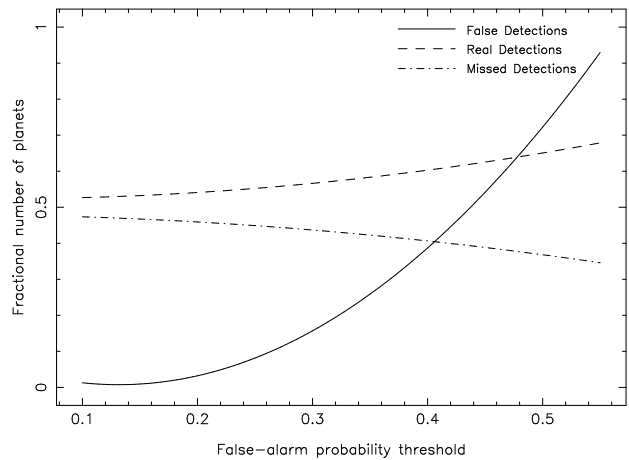}
  \caption{Dependence of false, real, and missed detections on the chosen
    periodic false-alarm probability threshold.}
\end{figure}

The range of real detection rates varies from around 55\% at a probability
threshold of 10\%, to around 65\% at a probability threshold of 50\%.
Considering that these numbers already take into account those planets
which fall below the theoretical accuracy curve discussed earlier, these
numbers may seem to indicate a quite low detection level. However, two
further factors which affect the number of detections from the data are the
number of data points acquired per target, and the period of the planetary
orbit relative to the observing window of the observations.

\begin{figure*}
  \begin{center}
    \begin{tabular}{cc}
      \includegraphics[angle=270,width=8.2cm]{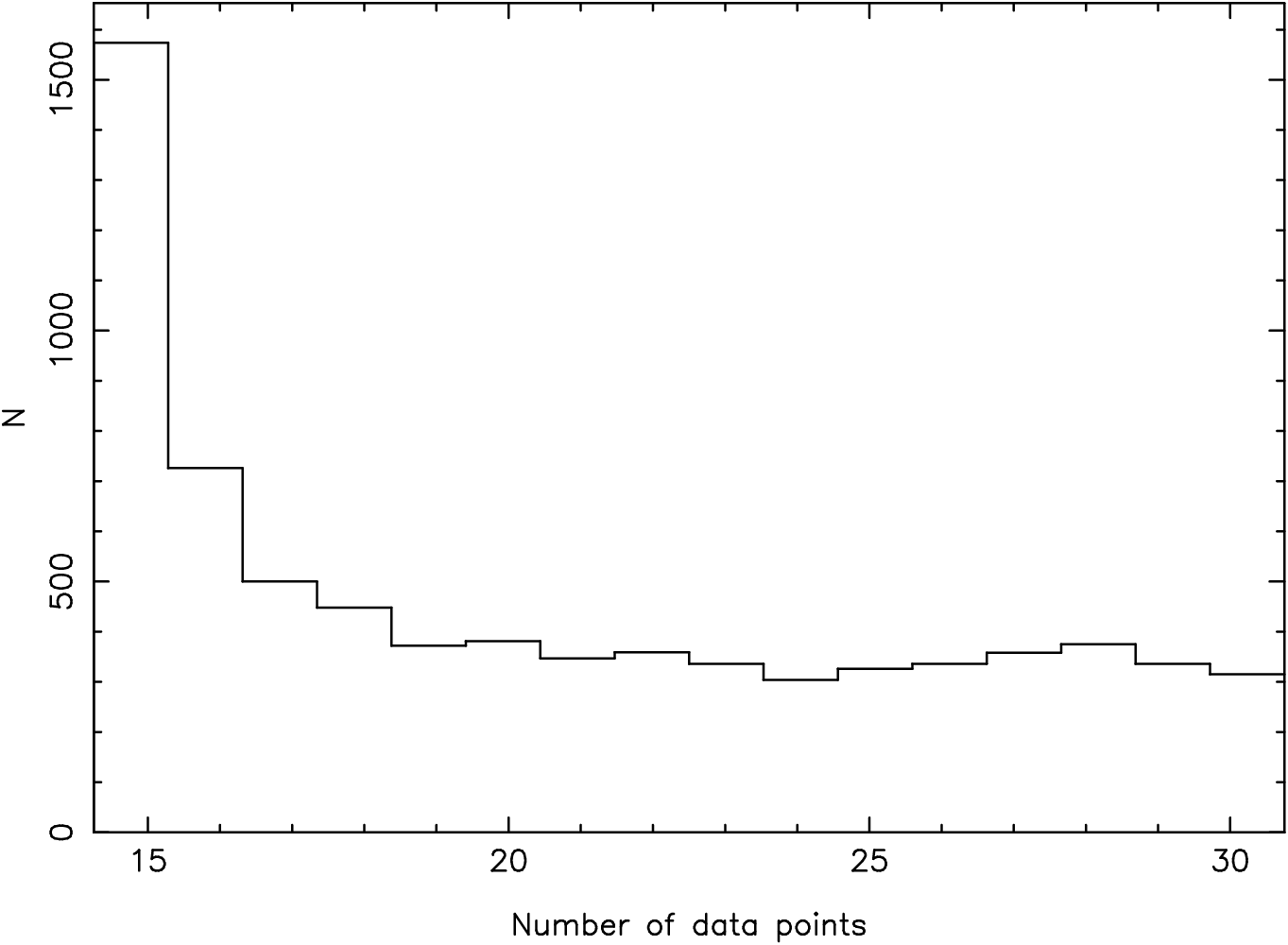} &
      \includegraphics[angle=270,width=8.2cm]{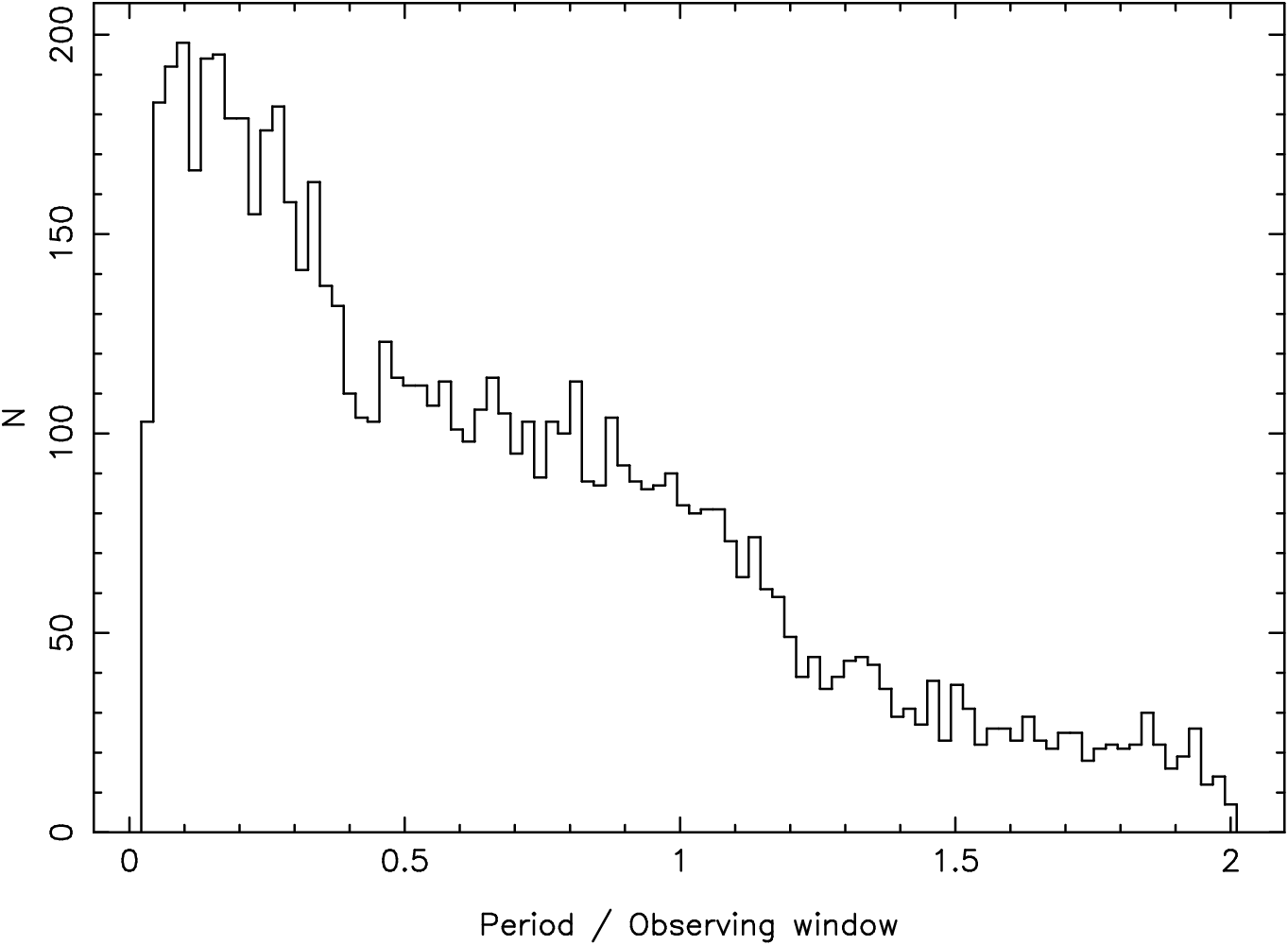} \\
    \end{tabular}
  \end{center}
  \caption{Histograms based on tests performed to determine the dependence of
  successful detections on the number of data points (left) and the planetary
  period (right).}
\end{figure*}

To provide quantitative estimates on how these factors effect the detection
rates, we performed several experiments using a simulated dataset in which
all of the stars are assumed to have planets and there are no interruptions
due to weather. This simulation was created to provide the most accurate
statistics since we are only interested in the number of successful detections,
not false detections, for these tests. A L-S periodogram analysis was performed
on each star's observations, and only those stars which satisfied the following
criteria were selected: (1) the probability of the periodic signal being a
false-alarm is less than 10\% and (2) the initial period estimate from the
periodogram is within 10\% of the true period. For each selected star the
number of data points and the period as a fraction of the observing window were
determined. Recall that the observing window for the simulated data is 30 days.

The resulting histograms are shown in Figure 8. The histogram of the number
of data points required for selection shows that the majority of stars need
$\sim 15$ data points to pass the test. The scatter in the data shown in the
histogram is a combination of two factors: (a) photon statistics dominating at
faint magnitudes and thus increasing the number of data points required for a
successful detection, and (b) periods longer than the observing window not
having sufficient period coverage to secure a detection. The second histogram
investigates the dependence of period (or phase) coverage on the number of
successful detections by expressing the period as a fractional duration of the
observing window. The histogram clearly shows an approximately linear
dependence with the number of datasets passing the test dropping to almost
zero once the period of the planetary orbit exceeds twice the length of the
observing window. This is a selection effect one would expect for detecting
planets with periods smaller than the observing window. However, it helps one
to understand the effect this factor has on the overall planet yield from a
duration limited survey when expressed in this fashion.

\subsection{The Fitting Algorithm}

There are many methods one can select when attempting to model astronomical
data, such as bayesian techniques, genetic algorithms, and simulated annealing.
Since the aim of this fitting algorithm is to provide a reasonably robust fit
in an automated fashion to a multitude of data, the method that was used for
this analysis is an iterative grid-search approach. The method is similar in
many respects to the amoeba, or downhill-simplex method, and proves to be
remarkably robust against falling victim to local minima. The iterative
grid-search method is described in more detail below.

The first step of the fitting process is to estimate a range values for each
fit parameters over which to conduct the search. Most of the inital guesses of
the fit parameters do not assume a priori knowledge per se, but rather rely on
the period provided by the L-S periodogram and the overall statistics of the
data itself. There are six parameters which are independently fit: the period
$P$, the systemic velocity $V_0$, the semi-amplitude $K$, the argument of
periastron $\omega$, the time at periastron $t_0$, and the eccentricity $e$.
The fitting of the argument of periastron and the eccentricity do use the a
priori knowledge that they are limited by $0 < \omega < 2 \pi$ and $0 < e < 1$
respectively.

After the minimum and maximum values for each of the fit parameters have been
established, a grid is constructed by dividing the range of values for each
parameter into equal segments. The number of segments the grid is divided
into depends upon the grid precision $g_p$ which is defined by the user. The
function (in this case, the radial velocity model described in section 3.3)
is then evaluated at $g_p + 1$ points on the grid for each parameter including
the current minimum and maximum values. The points on the grid which produce
the minimum $\chi^2$ are then used as the central values for the next iteration
which uses the grid precision to new minimum and maximum values over which to
search. This process is repeated until either the improvement in $\chi^2$ is
less than $10^{-3}$ or the number of iterations exceeds 15. The method is able
to provide excellent fits to almost all of the planetary data with little or
no human interaction.

Shown in Figure 9 are some typical examples of simulated radial velocity data
for stars harbouring a planet and the best-fit models. These fits only used a
grid precision of 6 and so coverged on the solution quite rapidly. The data have
a range of parameter values, particularly the period and the eccentricity. Since
the eccentricity defines the shape of the radial velocity curve, it tends to be
the most difficult parameter to obtain a reliable fit to, especially if there is
insufficient phase coverage such as planets with periods longer than the
observing window.

\begin{figure*}
  \begin{center}
    \begin{tabular}{cc}
      \includegraphics[angle=270,width=8.2cm]{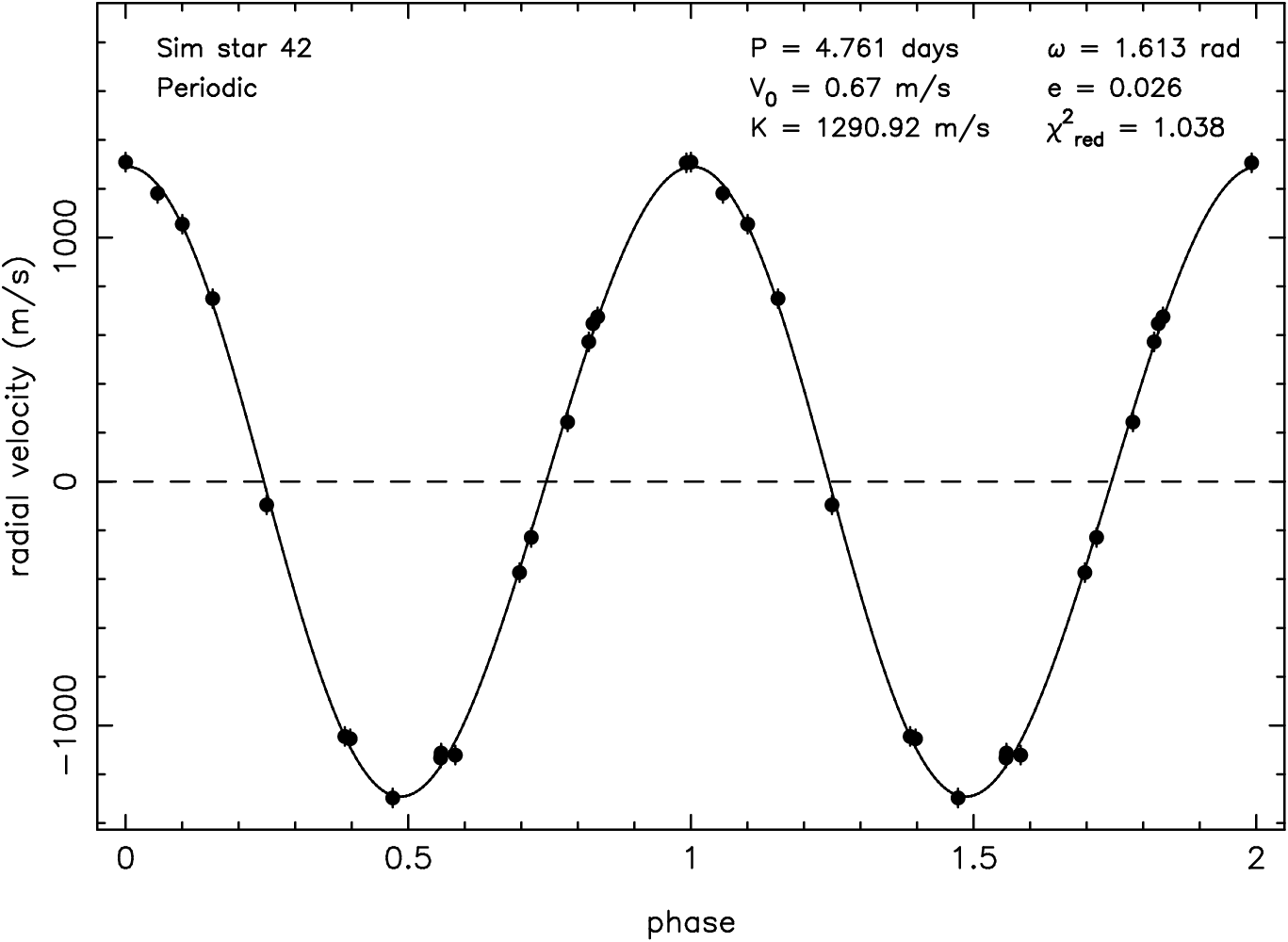} &
      \includegraphics[angle=270,width=8.2cm]{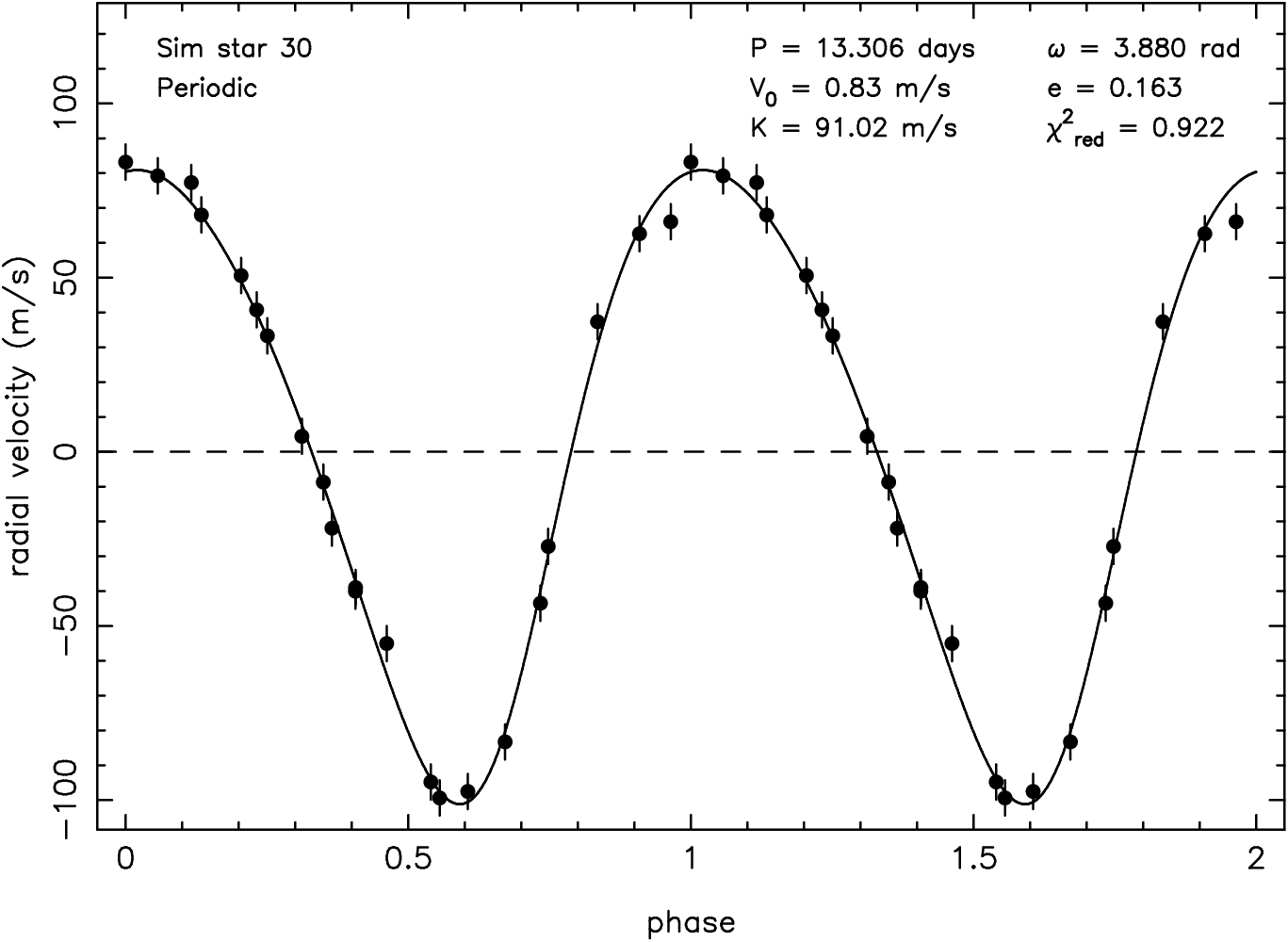} \\
      \includegraphics[angle=270,width=8.2cm]{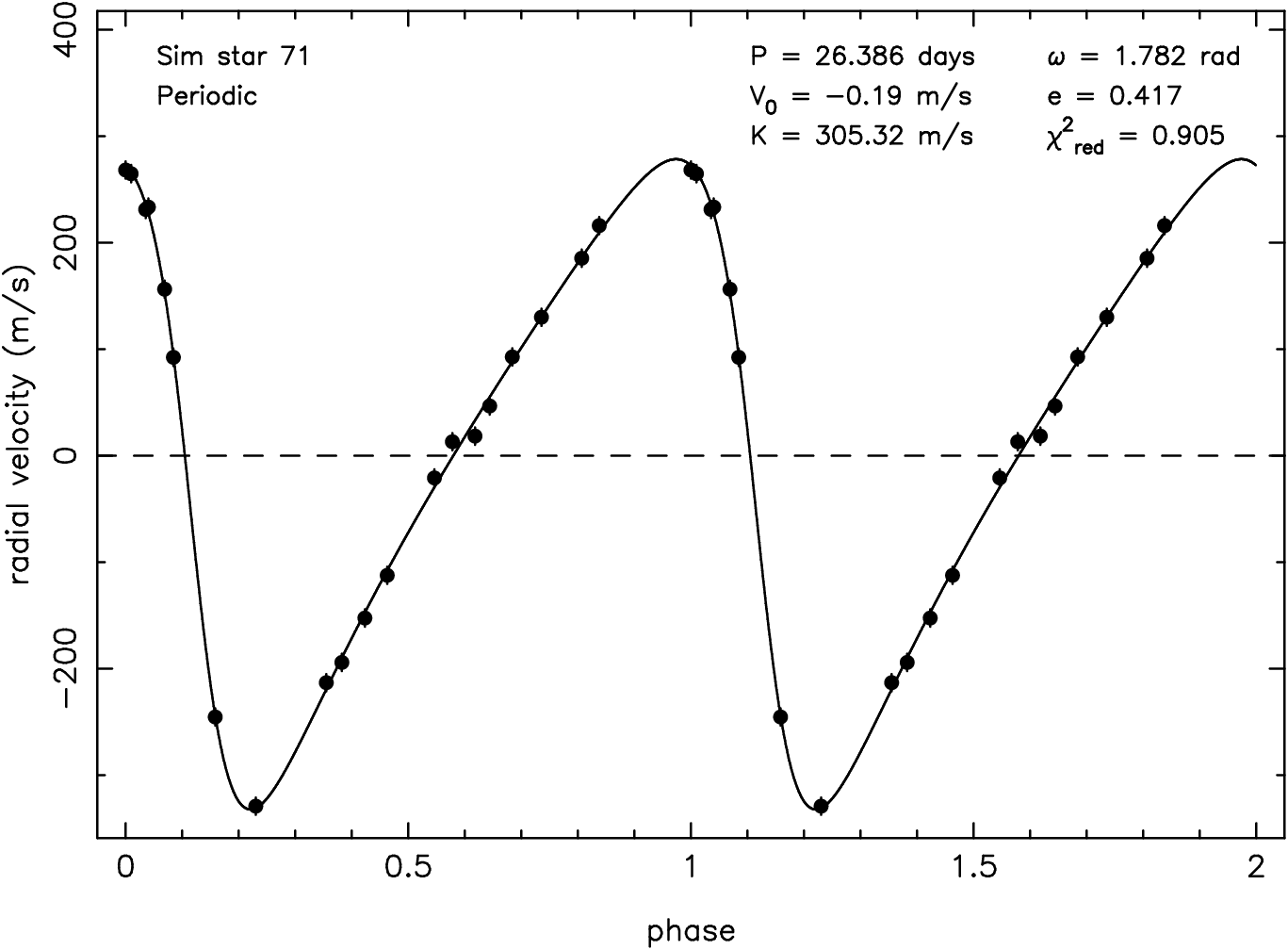} &
      \includegraphics[angle=270,width=8.2cm]{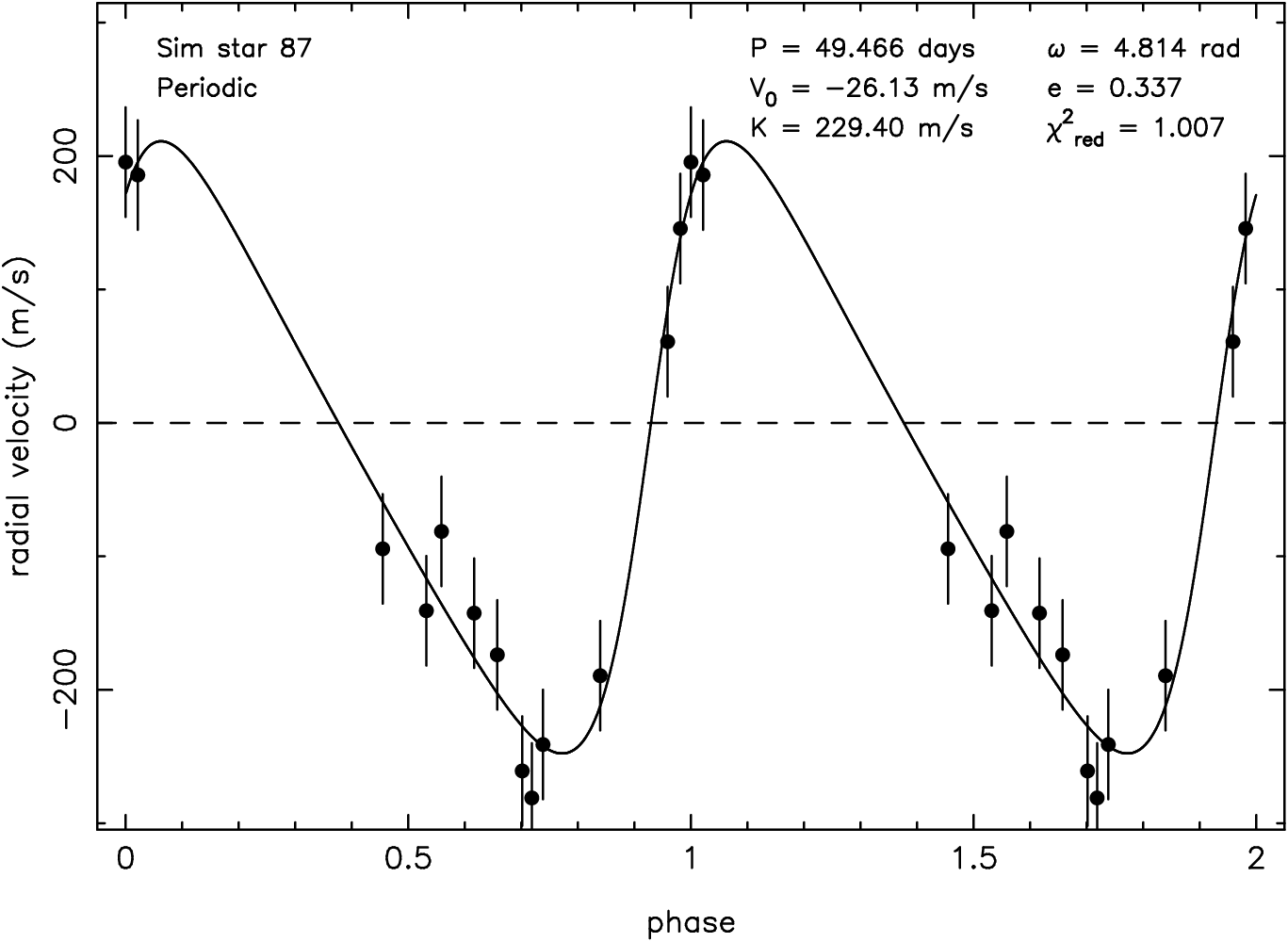} \\
    \end{tabular}
  \end{center}
  \caption{Examples of simulated radial velocity data for stars harbouring a
    planet and the best-fit models.}
\end{figure*}

The iterative grid-search does have several features which should be noted. The
grid precision essentially defines the resolution of the constructed grid. As
such, one might naively expect that a linear increase in the grid precision
will lead to a linear improvement in the fit of the model to the data. Although
this generally tends to leads to a better $\chi^2$, the relationship is rather
more complex and in fact one can only be assured of an improved fit if the
chosen $g_p$ is twice the previous value, where the previous value is an even
number. Even numbers tend to lead to better fits due to the way in which the
grid is re-created on each iteration, meaning that the current best fit will be
included as a grid-point in the refined grid. Since the number of model
evaluations for a given iteration is equal to $(g_p + 1)^N$ where $N$ is the
number of fit parameters, the process can quickly becomes computationally
expensive. It is therefore often the case that one uses a higher grid precision
as a means of refining a fit at a later stage. Lastly, the fit relies quite
heavily upon the inital guess of the planetary period and the mostly likely
cause of a scewed fit is a relatively strong alias in the periodogram which has
corrupted the fit. Provided one is able to recognise when this has occurred, it
is quite straightforward to solve.

\subsection{Uncertainties in Fit Parameters}

The calculation of the uncertainties on each of the fit parameters is performed
in a separate process to the model fitting described above. The uncertainties in
the parameters are estimated using a Monte-Carlo simulation which passes the
data through a gaussian filter based upon the error bars. Each successive
dataset is then passed through the fitting procedure resulting in an independent
set of values for the fit parameters which are then stored in arrays. Repeating
this operation a suitable number of times leads to a probability distribution of
values for each parameter from which one can compute the sample variance to
define the uncertainty in that parameter.

This is a powerful and robust method for defining the fit uncertainties but has
certain drawbacks which should be mentioned. This method assumes that the
uncertainties are gaussian in nature which is not always the case for $\omega$
and $e$. Indeed the probility distribution for $\omega$ is often double-peaked,
which one can overcome by considering the two peaks as seperate distributions.
This method is also computationally expensive depending on how large a
simulation is considered adequate to produce reliable statistics. However, since
the uncertainties are calculated in a separate process to the overall fitting,
this only need be performed when necessary.

\section{Target Selection}

Target selection is a crucial step to exclude giant stars and unstable stars from
an exoplanet radial velocity survey. This is normally achieved through the use of
catalogue information (such as that provided by Tycho-2 \citep{hog00} and 2MASS
\citep{skr06}), reduced proper motion diagrams, and various other techniques.
This process normally excludes the majority of stars in a given field as suitable
targets since giant stars tend to be dominant in a magnitude-limited survey.

\begin{table*}
\begin{minipage}{13cm}
  \caption{Number of dwarf stars available based on Besancon models of the Kepler
    field, COROT center field, and COROT anti-center field.}
  \begin{tabular}{@{}lrrrrrrr}
    Field & $V$ mag & stars & giants & dwarfs & \% dwarfs & dwarfs\sq\degr &
    Sloan FOV\\
    Kepler            & $< 14$ & 62114 & 43746 & 18368 & 29.6 & 174 & 1218\\
                      & $< 13$ & 25942 & 20843 &  5099 & 19.7 &  48 &  336\\
                      & $< 12$ & 10117 &  8701 &  1416 & 14.0 &  13 &   91\\
    COROT center      & $< 14$ & 15289 & 13017 &  2272 & 14.7 & 227 & 1589\\
                      & $< 13$ &  4803 &  4207 &   596 & 12.4 &  59 &  413\\
                      & $< 12$ &  1439 &  1298 &   141 &  9.8 &  14 &   98\\
    COROT anti-center & $< 14$ &  6570 &  4833 &  1737 & 26.4 & 173 & 1211\\
                      & $< 13$ &  2284 &  1821 &   463 & 20.3 &  46 &  322\\
                      & $< 12$ &   847 &   720 &   127 & 15.0 &  12 &   84\\
  \end{tabular}
\end{minipage}
\end{table*}

As a consquence of this, an important issue for multi-object surveys is having
enough suitable targets within the given FOV. To investigate this aspect in
detail, Besan\c{c}on models for the COROT center ($l = 37.0^\circ; b =
-6.9^\circ$) and anti-center ($l = 212.6^\circ; b = -1.1^\circ$) fields were
constructed to compliment the existing Kepler field simulation. The benefits of
choosing these fields are: (a) they are the subject of current and future
searches for transiting exoplanets making them likely targets for complimentary
radial velocity surveys, and (b) they represent diverse stellar densities due
to their very different locations. The size of each of these fields was chosen
to be 10 square degrees to provide a large enough sample, though the ``eye'' of
COROT is significantly larger than this. For each of these fields, the giant
stars were distinguished from the dwarf stars and a colour cutoff was imposed
which excluded stars earlier than $\sim$F7.

Table 1 shows the results of this study which estimates the number of suitable
targets per field. This was performed for a range of magnitude depths, with the
bright end set to $V = 6$. An additional column was produced which scales the
results to the size of the Sloan 2.5m FOV which is around 7 square degrees.
These results are consistent with target number estimates from catalogue
information, including the percentage rise in dwarf stars with increasing
magnitude depth. This places fairly tight constraints on the number of planned
fibres for a given instrument and magnitude depth, depending on the FOV. For
example, the Keck ET instrument is currently able to monitor $\sim 60$ targets
simultaneously \citep{ge06b}, from which these conservative estimates show that
there will be an adequate number of targets at a magnitude depth of $V = 12$ or
fainter.

\section{Expected Survey Results}

The results one can expect from a given survey depends upon a number of factors,
including the observing site, the efficiency of the instrument, and the number
of stars observed. Here we estimate the overall survey results for our particular
simulation and discuss how this can be translated into the expected number of
planet discoveries from a particular experiment.

\subsection{Total Planet Yield}

The total number of planets in the simulated sample is $\sim 750$. As shown in
Figure 4, the number of planets undetectable due to the photon-limit of the
survey is around 33\% of this number, or 250 planets. This leaves $\sim 500$
planets remaining in the sample. The number of planets lost during the sifting of
the data depends upon how strict a criteria is used to minimise the number of
false detections. If we adopt a false-alarm probability threshold of 10\%, thus
producing essentially zero false detections, then Figure 7 shows that 45\% of
the remaining planets will be lost, or 225 planets. Thus, the number of planets
detected in the survey is around 37\% of the total number of planets, or $\sim
275$ planets. Since there are around 25000 stars in the survey in this case,
this means a detection rate of around 1\% so that for every 100 stars surveyed,
1 star will have a planet that will be detected.

It must be remembered that this planet yield assumes a ``blind'' survey, meaning
there are no other selection criteria for the stars included in the survey for the
specified magnitude range other than to distinguish giant stars from dwarf stars.
This can be improved upon using techniques discussed in later sections. It is a
relatively straightforward calculation to scale this planet yield with the
magnitude depth of the survey based upon the results shown in Table 1. In
principle, each change in unit magnitude changes the volume of survey space and
therefore the integrated star counts by a factor of $\sim 4$. However, this is not
true in practice due to interstellar absorption and the direction dependent
spatial distribution of stars. For example, according to Table 1 the total number
of stars in the Kepler field increases by a factor of $\sim 2.5$ per unit increase
in magnitude depth for the range of magnitudes shown. For the COROT fields, this
factor is closer to $\sim 3.0$ due to the proximity of these fields to the
Galactic plane. The percentage number of dwarf stars increases with depth though
and so the change in dwarf stars for each change in unit magnitude is $\sim 3.5$.
If, for example, the magnitude depth of the survey is decreased from 14 to 13,
then the number of stars will decrease from 23708 to $\sim 6800$ and thus the
number of detected planets will decrease to $\sim 80$. Of course increasing the
magnitude depth of the survey requires an equivalent improvement in radial
velocity precision to prevent planetary signals becoming lost in the noise.

\subsection{Survey Duration Dependence}

If one increases the number of stars in the survey, then one can expect to
increase the overall planet yield by a proportional amount. This is not the case
for increasing the survey duration however since, although Figure 8 shows that
successful detections for a given period increases linearly with the observing
window, we have assumed a period distribution which is uniform in log space.
Thus, the change in the number of planets to which one is sensitive can be
determined by integrating over the region of the period distribution which
matches the survey duration. For example, if the survey duration is increased to
60 days, then the number of additional planets $N_a$ can then be approximately
estimated as follows
\begin{equation}
N_a = \frac{N_p(60)-N_p(30)}{N_p(60)} N_t
\end{equation}
where $N_p(30)$ and $N_p(60)$ are the number of planets with periods less than 30
days and 60 days respectively, and $N_t$ is the total number of detected planets.
For the simulation presented here, these numbers are 276, 344, and 275
respectively, resulting in an additional 67 possible planet detections at larger
periods.
As mentioned in section 3.4, this assumes that the planet distribution only
includes one planet per planet-harbouring star. The fractional number of
multiple exoplanetary systems is currently known to be $\sim 0.12$. Thus we
can also estimate the number of additional planets in longer orbits $N_m$ for
those systems already detected
\begin{equation}
N_m = 0.12 (N_t + N_a)
\end{equation}
leading to 41 additional planets in multiple-planet systems. This assumes that
the survey duration has been sufficiently extended to be sensitive to the outer
planets and that the observing strategy allows for continued monitoring of
those stars. This is of course a rough estimate, but is sufficient as a lower
limit as it is expected that the known fractional number of multiple exoplanetary
systems will only increase with time. It is also likely that these systems will
contain planets with longer periods that can be discovered through continued
monitoring.

\subsection{Eccentricity Bias}

The chosen observing schedule undoubtedly influences the sensitivity to various
planetary orbital parameters. Perhaps the most significant effect on the
detection efficiency is due to the eccentricity of the planet. Highly eccentric
orbits exhibit a rapid change in radial velocity which is likely to result in a
failure of their periodic detection via fourier analysis if there are
reasonably large gaps in the observing schedule. Since most of the observations
will occur when the planet is closest to apastron when the radial velocity
changes are at their lowest, the overall effect in terms of the detection will
be similar to that of planets whose period is much large than the duration of
the survey as discussed earlier. The highly eccentric planets can still be
detected if their apastron radial velocity variations exceed that required by
the variability test, such as the one discussed in section 5.1. This will be
true in most cases.

To provide a quantitative estimate of the bias against detecting eccentric
planets, a simulation was constructed identical to the one described in section
4 except for the following: (1) the eccentricity was uniformly distributed
between 0 and 1 and (2) the 30 days of observations were split into three groups
of 10 consecutive days with each group separated by 20 days. This observing
strategy is quite typical of experiments which have only intermittent access to
telescope time. This configuration was used to simulate over $4.7 \times 10^6$
planet-harbouring datasets which were passed through the sifting algorithm
described in section 5.1 and using a strict false-alarm probability threshold
to eliminate false detections.

\begin{figure}
  \includegraphics[angle=270,width=8.2cm]{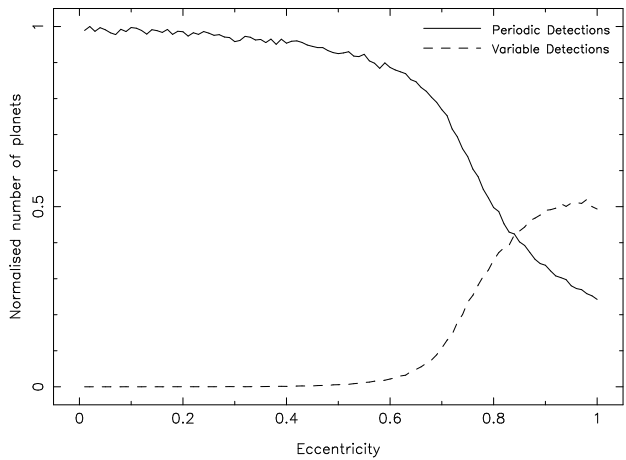}
  \caption{The effect of planetary eccentricity on the planet detection efficiency
    for an observing schedule which utilises a 10 consecutive nights for 3 months
    strategy.}
\end{figure}

The results of this exhaustive simulation are shown in Figure 10. The detection
efficiency begins to become severely affected for planets with eccentricities
greater than $\sim 0.6$, as shown by the solid line which represents datasets
flagged as periodic by the fourier analysis. If one includes the planets which
have been flagged as having an rms scatter significantly greater than the standard
deviation of the data, then one is able to recover many of the highly eccentric
planets despite the lack of phase coverage. This is an expected result but worth
highlighting in a qualitative fashion in order that surveys can judge their bias
against the eccentric planets when planning the observing strategy.

\subsection{Total Transiting Planets}

The popularity of the transit method to detect extra-solar planets is
self-evident by the large number of independent teams currently using the
technique for planet hunting. The reason that the transit method has
become so popular is due to the radial velocity surveys discovering a
relatively high number of ``hot Jupiters'' orbiting solar-type stars. In
fact, 0.5\%--1\% of Sun-like stars in the solar neighbourhood have been
found to harbour a Jupiter-mass companion in a 0.05 AU (3--5 day) orbit
\citep{lin03}. The probability, $P_t$, of a planet producing an observable
transit can be described by
\begin{equation}
P_t = \frac{(R_\star + R_p)}{a} \approx \frac{R_\star}{a}
\end{equation}
where $R_\star$ and $R_p$ are the radii of the star and planet respectively.
It is reasonable to assume that the orbital plane of these short-period
planets are randomly oriented, which according to the geometric transit
probability means that approximately 10\% of these planets will transit the
face of their parent star as seen by an observer. Thus, the transit method is
favoured considering the conclusion that close to 1 in 1000 solar-type stars
will produce detectable transits due to an extra-solar planet. Since this
transit method clearly favours large planets orbiting their parent stars at
small orbital radii, many of the hot Jupiters discovered via radial velocity
surveys are expected to also exhibit a photometric transit signature.

To approximate the number of transiting planets expected from the simulated
survey, the geometric transit probability was calculated for each star/planet
system as part of the Monte-Carlo simulation discussed in section 3.4. In
each case, it was randomly determined if the planet does indeed transit the
parent star. For the 751 planets contained in the simulation, it was
consistently found that 30 observable transiting planets can be expected to
be present. This estimation takes into account the entire period range which
is why it is significantly smaller than 10\% of the total planet population.
It is assumed that we will only detect transiting planets as a result of
photometrically monitoring the detected planets with shorter periods, and
hence the total number of detected transiting planets will be a factor of
37\% smaller, as per section 5.1, or 11 transiting planets. Observing
strategies for optimal photometric detection of the transiting planets from
radial velocity surveys are discussed elsewhere \citep{kan07}.

\section{Discussion}

One of the main goals of this paper has been to approximate the expected number
of planets from a large scale radial velocity survey. There are limitations to
the simulation which should be noted, however. Section 3.4 describes the method
used to create a realistic distribution of planet parameters from which the
characteristics of each planet was drawn from. Great effort has been made to
ensure that these parameter distributions are as free from bias as possible. One
of the most obvious cases where bias is possible to have crept in is the period
distribution. The period range was chosen to ensure that as many stars as
possible will have been monitored for at least as long as the upper end of the
chosen range. In addition, transiting planets were excluded due to their obvious
bias towards shorter periods. Even so, any unaccounted for bias towards shorter
periods will lead to an over-estimation in the number of planets detected
depending on the survey duration.

The method used for fitting the radial velocity data in this paper is described
as an iterative grid-search method. There are certainly other methods used
elsewhere for which a comparison would prove useful. A powerful method used in
a wide variety of astronomical applications is the method of Bayesian analysis.
Bayesian analysis is a statistical procedure which uses a prior distribution
of the parameter values in combination with the observed distribution to
create a probability map over all possible values. Thus Bayesian methods
deliver not only the the optimal parameter values but also their complete joint
posterior probability distribution which also determines the uncertainties in
the parameters. This method has been used successfully for extracting periodic
exoplanet signatures in radial velocity data \citep{for06,gre05}. Another
common data-fitting method is the use of genetic algorithm \citep{cha95}. Like
simulated annealing, this method is specifically designed to overcome the
difficult problem of avoiding local minima in $\chi^2$ space by allowing
individual groups of fits to develop is a manner similar to evolutionary
biology where the best traits of each fit are passed on to the next generation.
The main purpose of the current sifting and fitting code is to flag those data
for which periodic signals are suspected to be present from which any optimal
fitting method can then be adopted.

The Besan\c{c}on model provides a very effective and convenient method to aid
in selecting optimal fields in which to conduct a particular survey. As
mentioned earlier, this approach was used by \citet{jen05} to choose the best
field for the upcoming Kepler mission. In particular, both transit and radial
velocity surveys are concerned with the number of giant stars in the survey
fields as substantial numbers of these stellar types can drastically reduce
the expected planet yield. Creating comparative Besan\c{c}on models for a
selection of proposed field centers can thus be used in combination with
typical methods for sifting dwarf stars, such as reduced proper motion
diagrams, to provide optimal target selection for the survey.

An aspect which is not considered in the simulation is the inclusion of
high-mass stars in the sample, since most surveys to date have focused their
attention on F--G--K stars. It has been noted, for example by \citet{gal05},
that high-precision radial velocity measurements of early A--F stars is
exceptionally difficult, mostly due to the lack of spectral lines usable for
cross-cerrelation and the line-broadening caused by high rotational
velocities. For this reason, one may decide that early-type stars are
inaccessable to the survey in question. However, the results of the simulation
are not substantially changed if high-mass are excluded since they only
contribute a small percentage to the total number of stars.

The Besan\c{c}on model for the Kepler fields predicts a metallicity
distribution which increases towards sub-solar metallicity. This low
metallicity bias is not an unexpected result, since the metal-poor nature of
the field stars will always be true to varying degrees for a magnitude-limited
survey which essentially only probes the solar neighbourhood. Surveys such as
that performed by the N2K consortium \citep{rob06} estimate atmospheric
parameters via low-resolution spectroscopy to screen low-metallicity stars
from their survey targets. The survey proposed using the Keck ET instrument
will be able to screen giant stars and, in a more limited capactity, active
stars via photometry available from photometric catalogues, but is otherwise a
``blind'' survey in many regards. The survey is more than able to compensate
for this simply due to the vast amount of stars one is able to survey thanks
to the multi-object capabilities of the instrument. Hence, with an equivalent
radial velicity precision, one expects that N2K will detect a higher rate of
planets per survey star, but the Keck ET survey will detect a much higher
overall number of planets.

\section{Conclusions}

With the introduction of multi-object spectrographs capable of high-precision
measurements, the rate of planet detection is expected to rise significantly.
This paper presents simulations of a sample of stars from the Kepler field
which demonstrate the advantages of using such an instrument and a rough
guide on how many planets could be expected to be detected given a certain
noise model and magnitude limit.

The simulated data was derived from a Besan\c{c}on model of the Kepler field
which is an ideal location to perform a radial velocity survey complimentary
to planned future transit surveys. The stellar metallicities from the model
allowed the estimation of planet frequency based on a Monte-Carlo simulation
which utilises the well-known planet-metallicity correlation. This produced
751 planets from 23708 stars with a cumulative histogram which shows the
rapid decline in planet-harbouring stars beyond a probability of around 5\%.
The distribution of planetary parameters were derived from a numbers of
sources to produce as accurate a representation as possible of the expected
planets in the simulation.

The noise model and the simulated planets were the subject of an additional
Monte-Carlo simulation which demonstrated the number of planet detections
one can expect from a given radial velocity precision. This showed that, in
general, doubling the number of stars will increase the number of planets a
factor of five more than doubling the precision of the instrument. The number
of targets is thus by far the dominant factor in a survey, hence the
advantage of multi-object instruments is clear.

When the number of targets is increased significantly, the sifting of the
data for planetary signatures becomes an important process in order to reduce
the number of false detections and increase the number of real detections in
a manner which is not too computationally expensive. By using a weighted
Lomb-Scargle fourier analysis in combination with variability statistics, it
is shown that only a slight reduction in the number of real detections leads
to a complete elimination of false detections. Also, it is shown that at least
15 data points is generally needed for a significant detection. The number of
these detections decreases linearly with the fractional period as compared to
the observing window.

The results from the simulation showed that around 37\% of the total number of
planets in the survey could be feasably detected, leading to around 1 detected
planet for every 100 stars surveyed. This is dependant upon the magnitude
limit of the survey and the survey duration, an increase of which will lead to
not only longer period planets, but also multiple planetary systems. The
number of transiting planets was calculated from the geometric transit
probability and was found to be around 11 transiting planets over the entire
period range. This simulation was designed based upon the current survey being
performed using the Keck ET instrument, however these planet yields can be
quite easily scaled to take into account the specifics of any particular planet
survey.

Since most multi-object surveys are most sensitive to Jovian planets with
periods less than around 5 days, the total planet yield is quite sensitive to
the chosen period distribution. As the duration of existing surveys increase,
an unbiased knowledge of the period distribution will improve and will lead to
more accurate simulations. With many space missions, such as Kepler, which are
expected to discover many hundreds of extra-solar planets in the near future,
multi-object radial velocity surveys will provide invaluable complimetary
orbital parameters for these systems.

\section*{Acknowledgements}

The authors would like to thank Suvrath Mahadevan, Eric Ford, Curtis DeWitt,
and Roger Cohen for several useful discussions. This work is supported by the
W.M. Keck Foundation, the University of Florida, and National Science
Foundation grant AST-0451407.

\end{document}